\documentclass[10pt,journal,final,twocolumn]{IEEEtran}

\usepackage{graphics}
\usepackage{subfigure}
\usepackage{epstopdf}
\usepackage{epsfig}
\usepackage[cmex10]{amsmath}
\usepackage{amssymb}
\usepackage{times}
\usepackage{ntheorem}
\usepackage{pifont}
\usepackage{stfloats}
\usepackage{bm}
\usepackage{color}
\usepackage{makecell}
\usepackage{array}
\usepackage{multirow}
\usepackage{booktabs}

\begin{document}
%
\title{Edge Learning Based Collaborative Automatic Modulation Classification for Hierarchical Cognitive Radio Networks}

\author{\IEEEauthorblockN{
Peihao~Dong,~\IEEEmembership{Member,~IEEE}, Chaowei~He, Shen~Gao, Fuhui~Zhou,~\IEEEmembership{Senior Member,~IEEE},\\
and Qihui Wu,~\IEEEmembership{Fellow,~IEEE}
}

\vspace{-0.3cm}
\thanks{
	P. Dong is with the College of Electronic and Information Engineering, Nanjing University of Aeronautics and Astronautics, Nanjing 211106, China, and also with the National Mobile Communications Research Laboratory, Southeast University, Nanjing 211111, China (e-mail: phdong@nuaa.edu.cn).
	
	C. He, F. Zhou, and Q. Wu are with the College of Electronic and Information Engineering, Nanjing University of Aeronautics and Astronautics, Nanjing 211106, China (e-mail: hcw0110@nuaa.edu.cn; zhoufuhui@ieee.org; wuqihui2014@sina.com).
	
	S. Gao was with the National Mobile Communications Research Laboratory, Southeast University, Nanjing 211111, China (e-mail: gaoshen@seu.edu.cn).	
}
}

\IEEEtitleabstractindextext{%
\begin{abstract}
In hierarchical cognitive radio networks, edge or cloud servers utilize the data collected by edge devices for modulation classification, which, however, is faced with problems of the computation load, transmission overhead, and data privacy. In this article, an edge learning (EL) based framework jointly mobilizing the edge device and the edge server for intelligent co-inference is proposed to realize the collaborative automatic modulation classification (C-AMC) between them. A spectrum semantic compression neural network is designed for the edge device to compress the collected raw data into a compact semantic embedding that is then sent to the edge server via the wireless channel. On the edge server side, a modulation classification neural network combining the bidirectional long-short term memory and attention structures is elaborated to determine the modulation type from the noisy semantic embedding. The C-AMC framework decently balances the computation resources of both sides while avoiding the high transmission overhead and data privacy leakage. Both the offline and online training procedures of the C-AMC framework are elaborated. The compression strategy of the C-AMC framework is also developed to further facilitate the deployment, especially for the resource-constrained edge device. Simulation results show the superiority of the EL-based C-AMC framework in terms of the classification accuracy, computational complexity, and the data compression rate as well as reveal useful insights paving the practical implementation.
\end{abstract}

}

\maketitle

\IEEEdisplaynontitleabstractindextext

%
\IEEEpeerreviewmaketitle

\section{Introduction}

\IEEEPARstart{D}{espite} the skyrocketing development, wireless networks are faced with unprecedented challenges in terms of the spectrum scarcity and transmission security. Automatic modulation classification (AMC) is anticipated to play a crucial role in constructing effective solutions since it endows the intended receiver or monitor with the ability of recognizing the signal modulation type with the minimal prior information \cite{O. A. Dobre}. AMC was born for the military purpose, and was then widespreadly applied to the civilian wireless networks in past decades, such as the spectrum surveillance, intelligent modem design, and malicious attack identification \cite{P. Panagiotou}--\cite{B. Dong}.

The AMC algorithms are mainly categorized into two branches: likelihood-based (LB) and feature-based (FB) approaches. LB approaches calculate the likelihoods of all candidate modulation schemes with respect to the received signal, and select that with the maximal likelihood \cite{P. Panagiotou}, \cite{J. L. Xu}. Despite the optimality in the Bayesian sense, the LB approaches require the accurate knowledge of channel characteristics and high computational complexity hindering the practical implementation. This problem can be well treated by the FB approaches, which extract the received signal features, e.g., cyclic moments, wavelet-based features, high-order cumulants, in a low-complexity manner for classification decision while possessing the near-optimal performance \cite{W. A. Gardner}--\cite{L. Han}. However, due to the nature of relying on manually extracted features, traditional FB approaches may be overwhelmed by the increasingly complicated wireless environments as well as the soaring number of signal emitters, which invokes the more scalable tool for feature extraction.

The revival of deep learning (DL) invigorates the FB approaches and the marriage of them tends to become the mainstream solution for the modern AMC since the deep neural network (DNN) can act as a versatile feature extractor \cite{T. J. O'Shea}. In \cite{T. J. O'Shea_b}, one of the works spearheading this direction, the DL-based AMC approach was proved feasible and superior under the real propagation environment inclusive of the effects of some key system parameters. In \cite{S. Peng}, convolutional neural network (CNN) was exploited for AMC and the data format of the received signals matching best with the proposed CNN structure was revealed. A modified generative adversarial network was designed to improve the classification accuracy by augmenting the training set composed of contour stellar images in \cite{B. Tang}. In \cite{F. Meng}, an end-to-end CNN architecture using the two-step training was developed for AMC with the enhanced generalization ability. Considering different recognition difficulties of the modulation types, a DL structure consisted of two concatenate CNNs with respective recognition objects was developed in \cite{Y. Wang}. The direction of DL-based AMC became further prosperous as various subsequent studies mushroomed focusing on feature extraction enhancement \cite{T. Huynh-The}--\cite{WangTWC2021}, lightweight design \cite{Y. Lin}--\cite{L. Guo}, few-shot learning \cite{L. Li}--\cite{W. Deng}, distributed framework \cite{B. Dong, P. Qi, WangTCCN2022}, adversarial defense \cite{S. Zhang}, and so on.

As the wireless networks are increasingly expanded and complicated, the wide-area spectrum surveillance and management become pretty necessary considering both the spectrum scarcity and transmission security. As a result, the traditional cognitive radio (CR) system will evolve into a hierarchical network consists of the edge device, edge server and/or cloud server \cite{P. Dong}. In this architecture, the spectrum semantics need to be known by the edge/cloud server for the global spectrum management or decision-making while the edge device mainly takes on the task of sensing data. In other words, the data sensing and spectrum cognition are conducted at different locations in the network. The signal modulation type can be regarded as a kind of spectrum semantics extracted from the spectrum data while the existing DL models for AMC cannot be mechanically applied in the hierarchical CR network. In \cite{S. Rajendran}, a long-short term memory (LSTM) network was proposed to enable the AMC at distributed low-cost sensors. However, the manner of directly reporting the classification result brudens the resource-constrained edge device with the whole computation load and may leak the secret information. On the other hand, if the DNNs for classification are deployed at the edge server, the edge device has to deliver the sensed data via the wireless channel, incurring the high communication overhead and the risk of exposing the raw data privacy. Edge learning (EL), a paradigm enabling the more flexible deployment of DL models at the edge \cite{W. Xu}, provides the potential solution for this problem. That is, the DNN can be split into two parts in order to be respectively deployed at the edge device and edge server, aiming to balance the computation load, reduce the communication overhead, and improve the safety simultaneously \cite{P. Dong}. In \cite{J. Park}, the idea of DNN model splitting was adopted for AMC by partitioning a residual network, where the in-phase/quadrature (I/Q) samples are first processed by the model segment at the device and then the cut-layer representation is passed to the remaining model segment at the server for modulation classification. Although the cut-layer representation hides the private information, its high-dimension requires a considerable transmission overhead. In addition, the performance relies on a large set of I/Q samples, burdening the edge device in the data sensing and processing phases.

To address the mentioned-above problems, in this article, we propose an EL-based collaborative automatic modulation classification (C-AMC) framework consisted of a spectrum semantics compression neural network (SSCNet) and a modulation classification neural network (MCNet) deployed at the edge device and edge server, respectively, by treating the signal modulation type as a kind of spectrum semantics. The main novelty and contribution can be summarized as follows:

\begin{itemize}[\IEEEsetlabelwidth{Z}]

\item[1)] SSCNet and MCNet are well designed to achieve the goal of the C-AMC framework in terms of balancing the computation load, reducing the transmission overhead, and guaranteeing the information security simultaneously. Specifically, a quite lightweight structure accommodating the resource-constrained edge device is designed for SSCNet to compress the collected raw data into a compact semantic embedding. Thanks to the low dimension of the semantic embedding and the more powerful computation capability at the edge server, MCNet incorporates the bidirectional long-short term memory (Bi-LSTM) and multi-head attention structures to achieve the high classification accuracy via the sufficient feature extraction.

\item[2)] The offline training procedure of the C-AMC framework is elaborated along with the insight on the generalization capability. A simple online training procedure is proposed by considering the update of SSCNet and MCNet over the air. The combination of offline and online training enables the C-AMC framework to adapt to a new scenario fast.

\item[3)] The compression strategy of the C-AMC framework is developed to further facilitate the deployment, especially for the resource-constrained edge device. The magnitude-based importance for a weight is analyzed, based on which the weight pruning procedure is elaborated. Then the post-training quantization is applied to further accelerate the model inference and reduce the model size. A layer-by-layer complexity analysis of the C-AMC framework is also provided.

\item[4)] Extensive simulation results are provided to show the superiority of the EL-based C-AMC framework over baseline schemes in terms of the classification accuracy, computational complexity, and data compression rate. Useful insights are extracted from the results to shed light on the practical implementation of the C-AMC framework.
\end{itemize}

By using the C-AMC framework, the following benefits can be gained for the hierarchical CR system: 1) The computation load for classification is allocated between the resource-constrained edge device and the edge server more properly so that the endurance of the edge device can be improved. 2) By transmitting the compact and intricate semantic embedding instead of the raw data or the classification result from the edge device to the edge server, the transmission overhead is reduced while the system safety is enhanced. 3) The framework is scalable and provides a paradigm for the hierarchical CR system to realize the cognition of various spectrum semantics besides the signal modulation type considered in this paper.

The rest of this paper is organized as follows. Section II introduces the basic signal model, based on which the EL-based C-AMC framework including SSCNet and MCNet is developed in Section III. Section IV elaborates the offline and online training procedures of the C-AMC framework. Section V developes the weight pruning and quantization based compression strategy for the C-AMC framework along with the complexity analysis. Simulation results are presented in Section VI, followed by Section VII giving concluding remarks.

\section{System Model}

In this section, the hierarchical CR network model is elaborated by considering the modulation classification as the cognitive task.

\begin{figure}[t]
	\centering
	\includegraphics[width=2.9in]{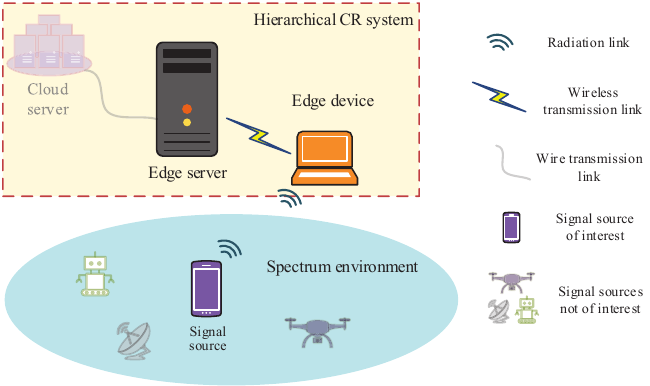}
	\caption{Hierarchical CR network model.}\label{HCR_model}
\end{figure}

As illustrated in Fig.~\ref{HCR_model}, consider a hierarchical CR network consisted of an edge device and an edge server aiming to recognize the modulation type of a source signal in the spectrum environment.\footnote{Since the limited capacity of the wireless link from the edge device to the edge server hinders the accurate global spectrum cognition for the hierarchical CR network, we focus on addressing this crux along with the computation load and transmission security problems. The transmission from the edge server to the cloud server can be carried out via the wire link and thus is not considered.} The cognitive process includes two phases, data sensing at the edge device and information transmission from the edge device to the edge server, along with the corresponding signal processing procedures.

\begin{figure*}[t]
	\centering
	\includegraphics[width=4.9in]{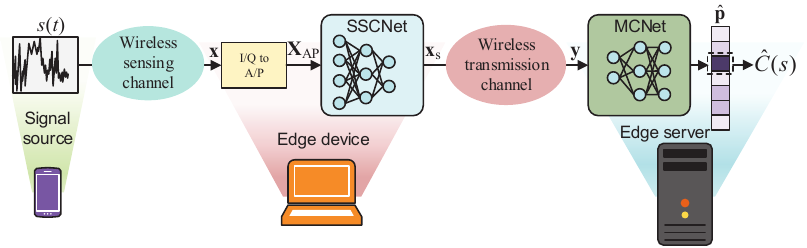}
	\caption{EL-based C-AMC framework.}\label{C_AMC}
\end{figure*}

\emph{Phase I (Data Sensing at the Edge Device):} The edge device keeps sounding the source signal $s(t)$ within $L$ sampling time instants and the received signal is given by
\begin{eqnarray}
	\label{eqn_xl}
	x[l] = h[l]e^{-j2\pi(\nu lT_{\mathrm{s}}+\vartheta)}s[l] + z[l],\quad l=1,\ldots,L,
\end{eqnarray}
where $h[l]$, $s[l]$, and $z[l]$ respectively denote the channel gain, modulated source signal, and additive white Gaussian noise (AWGN) at the $l$th sampling time instant, and $T_{\mathrm{s}}$, $\nu$, and $\vartheta$ represent the sampling period, frequency shift, and phase shift, respectively. The modulation type of the source signal, $C(s)$, is taken from the candidate set $\mathcal{M}$ including $M$ elements. Denote $\mathbf{x}=[x[1],\ldots,x[L]]$ as the vector form of the received signal. Then $\mathbf{x}$ is processed at the edge device to yield an $N$-dimensional vector expressed as
\begin{eqnarray}
	\label{eqn_xs_vec}
	\mathbf{x}_{\textrm{s}}=\phi(\mathbf{x}),
\end{eqnarray}
where $\phi(\cdot)$ represents the general mapping function of the signal processing. Specifically, $\mathbf{x}_{\textrm{s}}$ will give the classification result if $\phi(\cdot)$ is the modulation classification mapping while $\mathbf{x}_{\textrm{s}}=\mathbf{x}$ holds if $\phi(\cdot)$ is the identity mapping.

\emph{Phase II (Information Transmission from the Edge Device to the Edge Server):} In this phase, the edge device transmits $\mathbf{x}_{\textrm{s}}$ to the edge server via a certain air interface. The received signal after equalization at the edge server is expressed as
\begin{eqnarray}
	\label{eqn_y_elem_wise}
	y_{i}={x}_{\textrm{s},i}+w_i,\quad i=1,\ldots,N,
\end{eqnarray}
where ${x}_{\textrm{s},i}$ is the $i$th element of $\mathbf{x}_{\textrm{s}}$ and $w_i$ denotes the corresponding effective noise compounding the interference and AWGN. Then the edge server conduct the processing, $\varphi(\cdot)$, on $\mathbf{y}=[y_{1},\ldots,y_{N}]$ to recognize the modulation type of the source signal. Without loss of generality, consider the noise-free case, yielding
\begin{eqnarray}
	\label{eqn_c}
	\hat{\mathbf{p}}=\varphi(\mathbf{y})=\varphi(\phi(\mathbf{x})),
\end{eqnarray}
where $\hat{\mathbf{p}}\in \mathbb{R}^{M}$ is a probability vector indicating the most likely modulation type of $s(t)$ according to the index of its largest entry.
From (\ref{eqn_c}), $\phi(\cdot)$ and $\varphi(\cdot)$ collaborate to finish the modulation classification task in a complementary manner. For the specific cases mentioned above, $\varphi(\cdot)$ can be the identity mapping if $\phi(\cdot)$ is the modulation classification mapping and vice versa, which, however, aggravates the computation load for the edge device or incurs the high transmission overhead, in addition to the risk of exposing the secure information. Therefore, we aim to address these problems by elaborating $\phi(\cdot)$ and $\varphi(\cdot)$ in the following sections.

\section{EL-Based C-AMC Framework}

In this section, the EL-empowered C-AMC framework is constructed. The framework is overviewed first, followed by detailing its two component DL models, i.e., SSCNet and MCNet.

\subsection{Overview of Framework}

Inspired by EL, the C-AMC framework is able to balance the computation load of modulation classification between the edge device and edge server by respectively deploying SSCNet and MCNet for them, as illustrated in Fig.~\ref{C_AMC}. SSCNet compresses the sensed data into the low-dimensional spectrum semantic embedding for transmission by extracting features related to the modulation type therein. At the other end of the wireless channel, MCNet utilizes the noisy semantic embedding to predict the modulation type of the source signal. By doing this, a part of computation task of modulation classification can be offloaded from the edge device with the limited computing resource to the edge server. Besides, the low-dimensional semantic embedding reduces the transmission overhead and is difficult to decode for the potential eavesdroppers in the wireless channel.

Before being processed by SSCNet at the edge device, the sensed data $\mathbf{x}$ in the I/Q form is converted to the amplitude/phase (A/P) form as
\begin{eqnarray}
	\label{eqn_AP}
	\mathbf{X}_{\textrm{AP}}=\mathcal{P}(\mathbf{x})=\left[\begin{array}{cc}|x[1]|,\ldots,|x[L]|\\ \theta(x[1]),\ldots,\theta(x[L])\end{array}\right]^{T},
\end{eqnarray}
where $|x|$ and $\theta(x)=\mathrm{arctan}\frac{\mathrm{Im}(x)}{\mathrm{Re}(x)}$ respectively denote the amplitude and phase of $x$. It is noted that this preprocessing step helps SSCNet better extract the useful features. Then $\mathbf{x}_{\textrm{AP}}$ is processed by SSCNet to generate the spectrum semantic embedding with the compression rate $r=\frac{2L}{N}$, that is,
\begin{eqnarray}
	\label{eqn_xs_vec_SSCNet}
	\mathbf{x}_{\textrm{s}}=f(\mathbf{X}_{\textrm{AP}};\boldsymbol{\Theta}),
\end{eqnarray}
where $f(\cdot)$ denotes the mapping function of SSCNet parameterized by the weight set $\boldsymbol{\Theta}$. So the general mapping function $\phi(\cdot)$ can be represented as the composite of $\mathcal{P}(\cdot)$ and $f(\cdot)$, i.e., $\phi=f\circ \mathcal{P}$.

At the edge server, the noisy semantic embedding $\mathbf{y}=\mathbf{x}_{\textrm{s}}+\mathbf{w}$ with $\mathbf{w}=[w_{1},\ldots,w_{N}]$ is processed by MCNet to predict the modulation type as
\begin{eqnarray}
	\label{eqn_c_MCNet}
	\hat{\mathbf{p}}=g(\mathbf{y};\boldsymbol{\Phi})=g\left(f\left(\mathcal{P}(\mathbf{x});\boldsymbol{\Theta}\right)+\mathbf{w};\boldsymbol{\Phi}\right),
\end{eqnarray}
where $g(\cdot)$ denotes the mapping function of MCNet parameterized by the weight set $\boldsymbol{\Phi}$. Then the general mapping function $\varphi(\cdot)$ is instantiated as $g(\cdot)$, i.e., $\varphi=g$. Since $\mathcal{P}(\cdot)$ is a fixed operation, the design of $\phi(\cdot)$ and $\varphi(\cdot)$ becomes the design of SSCNet and MCNet represented by $f(\cdot)$ and $g(\cdot)$, respectively.

\subsection{SSCNet}

In the C-AMC framework, the role of SSCNet is to compress the sensed data with a high dimension into the compact semantic embedding using a lightweight architecture to accommodate the resource-limited edge device. To sufficiently compress the sensed data, the temporal correlation therein can be exploited by the powerful LSTM structure \cite{S. Rajendran}. Generally, more than one LSTM layer is needed to fully extract the temporal correlation, which, however, causes the high computation load for the edge device. To address this problem, we use the one-dimensional convolution (Conv1D) as a lightweight alternative to tentatively handle the temporal correlation instead of stacking multiple LSTM layers.

\begin{figure}[!t]
\centering
\subfigure[\scriptsize SSCNet]{\includegraphics[width=2.7in]{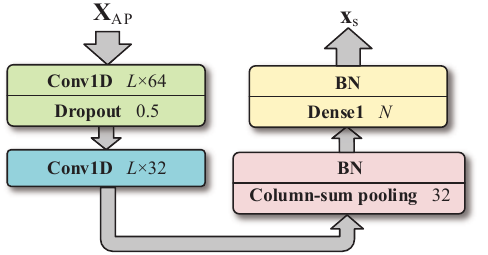}\label{SSCNet}}
  \vspace{0.1cm}
\subfigure[\scriptsize MCNet]{\includegraphics[width=3.4in]{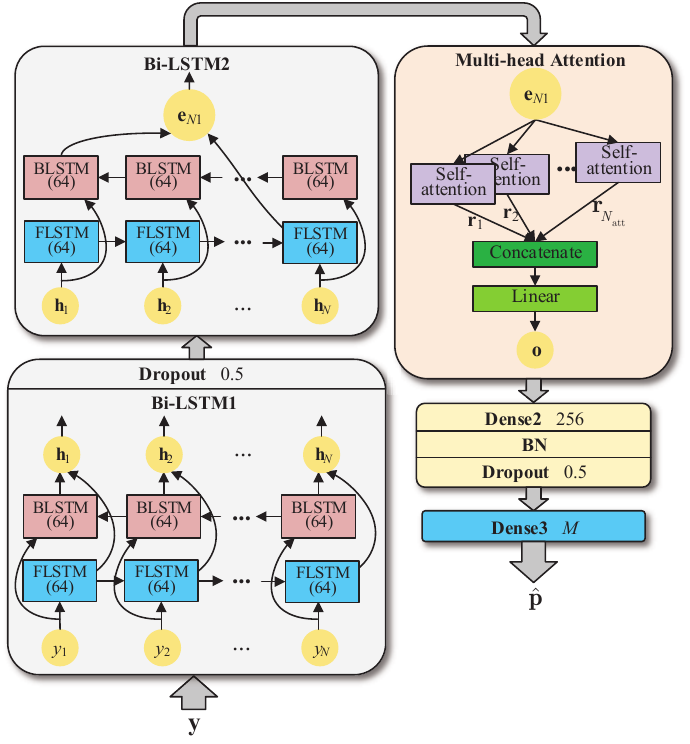}\label{MCNet}}
\caption{Architecture of SSCNet and MCNet.}\label{SSCNet&MCNet}
  \vspace{-0.6cm}
\end{figure}

Fig.~\ref{SSCNet} shows the detailed architecture of SSCNet. A Conv1D layer using $64$ kernels with length $8$ and rectified linear unit (ReLU) activation function is utilized first to filter the input $\mathbf{x}_{\textrm{AP}}\in \mathbb{R}^{L \times 2}$, in order to extract the local temporal correlation therein, after which a dropout layer with the dropout rate of $0.5$ is added to prevent overfitting and to improve the robustness of neurons. Then the $L \times 64$ feature map is processed by another Conv1D layer using $32$ kernels with length $8$ and ReLU to further distill the desired high-order spectrum feature, after which the size of the feature map is compressed from the $L \times 64$ to $L \times 32$. However, since the sequence length of the sampling data, $L$, usually ranges from several hundred to over a thousand, the feature map with size $L \times 32$ still consumes too much transmission overhead and thus needs to be further compressed. To this end, the column-sum pooling operation is applied to sum the feature map in a column-wise manner so that the feature map size can be dramatically reduced to $1 \times 32$.\footnote{As the $L \times 32$ feature map before pooling actually represents $32$ $L$-dimensional feature vectors, the column-sum pooling operation can reserve the feature diversity and indeed achieves the better performance compared with other pooling ways according to simulation trials.} Then a batch normalization (BN) layer is appended. Dense1 layer transforms the feature map to an $N$-dimensional vector. The scaled exponential linear unit (SELU) activation function is applied to Dense1 layer to ensure the neuron activity regardless of the sign of the input, that is,
\begin{eqnarray}
	\label{eqn_SELU}
	\mathrm{SELU}(x)=\delta \begin{cases}x, & x>0 \\ \alpha\left(e^x-1\right), & x \leq 0\end{cases}
\end{eqnarray}
with $\alpha>0$ and the scaling factor $\delta>1$. After BN, the $N$-dimensional real-valued specturm semantic embedding, $\mathbf{x}_{\textrm{s}}$, with $N \ll 2L$ is obtained for low-overhead transmission.

\subsection{MCNet}

MCNet should be designed to coordinate with SSCNet so that they can be integrated across the wireless channel to output the predicted modulation type at the edge server. Thanks to the much lower dimension of the noisy semantic embedding, $\mathbf{y}$, the powerful LSTM and attention layers can be applied in MCNet to fully extract the desired feature from $\mathbf{y}$ as well as to remove the noise imposed by the wireless channel by exploiting the global correlation therein while avoiding the high computational complexity.

Fig.~\ref{MCNet} shows the detailed architecture of MCNet. Considering the mutual dependence of the elements in $\mathbf{y}$, the Bi-LSTM structure is applied to extract more useful information. Bi-LSTM1 layer includes $N$ forward LSTM units and $N$ backward LSTM units, and the processing procedure can be expressed as
\begin{eqnarray}
	\label{eqn_hF}
	\mathbf{h}_{\mathrm{F},i}=\mathrm{FLSTM}(y_i;\mathbf{h}_{\mathrm{F},i-1},\mathbf{c}_{\mathrm{F},i-1},\boldsymbol{\Xi}_{\mathrm{F}}),
\end{eqnarray}
\begin{eqnarray}
	\label{eqn_hB}
	\mathbf{h}_{\mathrm{B},i}=\mathrm{BLSTM}(y_i;\mathbf{h}_{\mathrm{B},i+1},\mathbf{c}_{\mathrm{B},i+1},\boldsymbol{\Xi}_{\mathrm{B}}),
\end{eqnarray}
\begin{eqnarray}
	\label{eqn_hi}
	\mathbf{h}_{i}=[\mathbf{h}_{\mathrm{F},i}, \mathbf{h}_{\mathrm{B},i}],\quad i=1,\ldots,N,
\end{eqnarray}
\begin{eqnarray}
	\label{eqn_H}
	\mathbf{H}=[\mathbf{h}_{1}^{T},\ldots,\mathbf{h}_{N}^{T}]^{T},
\end{eqnarray}
where $\mathbf{h}_{\mathrm{F},i-1}$ and $\mathbf{c}_{\mathrm{F},i-1}$ denote vectors passed from the previous forward LSTM unit with the parameter set $\boldsymbol{\Xi}_{\mathrm{F}}$ including all the weight matrices and bias vectors of the unit, and $\mathbf{h}_{\mathrm{B},i+1}$ and $\mathbf{c}_{\mathrm{B},i+1}$ denote vectors passed from the previous backward LSTM unit with the parameter set $\boldsymbol{\Xi}_{\mathrm{B}}$ including all the weight matrices and bias vectors of the unit. The $i$th pair of forward and backward LSTM units respectively transform $y_i$ into the vectors $\mathbf{h}_{\mathrm{F},i}\in \mathbb{R}^{64}$ and $\mathbf{h}_{\mathrm{B},i}\in \mathbb{R}^{64}$, which are then concatenated to yield $\mathbf{h}_{i}$ as the output of this pair. Stacking $\mathbf{h}_{1},\ldots,\mathbf{h}_{N}$ gives the output of of Bi-LSTM1 layer, i.e., $\mathbf{H}\in \mathbb{R}^{N\times 128}$. After Bi-LSTM1 layer, a dropout layer with the dropout rate of $0.5$ is appended. Bi-LSTM2 layer also consists of $N$ forward LSTM units and $N$ backward LSTM units while it only passes the output of the final units of the two directions to the following layer for processing, that is,
\begin{eqnarray}
	\label{eqn_qF_N}
	\mathbf{e}_{\mathrm{F},N}=\mathrm{FLSTM}(\mathbf{h}_{N};\mathbf{e}_{\mathrm{F},N-1},\mathbf{d}_{\mathrm{F},N-1},\boldsymbol{\Pi}_{\mathrm{F}}),
\end{eqnarray}
\begin{eqnarray}
	\label{eqn_qB_1}
	\mathbf{e}_{\mathrm{B},1}=\mathrm{BLSTM}(\mathbf{h}_{1};\mathbf{e}_{\mathrm{B},2},\mathbf{d}_{\mathrm{B},2},\boldsymbol{\Pi}_{\mathrm{B}}),
\end{eqnarray}
\begin{eqnarray}
	\label{eqn_qN1}
	\mathbf{e}_{N1}=[\mathbf{e}_{\mathrm{F},N}, \mathbf{e}_{\mathrm{B},1}],
\end{eqnarray}
where the notations are similar to Bi-LSTM1 layer. Since the units of two directions in Bi-LSTM2 layer respectively transform $\mathbf{h}_{i}\in \mathbb{R}^{128}$ to $\mathbf{e}_{\mathrm{F},i}\in \mathbb{R}^{64}$ and $\mathbf{e}_{\mathrm{B},i}\in \mathbb{R}^{64}$, the size of the feature map output by this layer, $\mathbf{e}_{N1}$, is $1\times 128$. Then $\mathbf{e}_{N1}$ is processed by an $N_{\mathrm{A}}$-head self-attention layer. The output of the $n$th self-attention head is given by
\setlength{\arraycolsep}{0.1em}
\begin{eqnarray}
\label{eqn_atten_n} \mathbf{r}_{n}=\mathrm{S\text{-}Att}(\mathbf{e}_{N1}\mathbf{W}_{\mathrm{Q}}^{(n)},\mathbf{e}_{N1}\mathbf{W}_{\mathrm{K}}^{(n)},\mathbf{e}_{N1}\mathbf{W}_{\mathrm{V}}^{(n)}),
\end{eqnarray}
where $\mathbf{W}_{\mathrm{Q}}^{(n)},\mathbf{W}_{\mathrm{K}}^{(n)},\mathbf{W}_{\mathrm{V}}^{(n)}\in \mathbb{R}^{128\times d_{\mathrm{A}}}$ denote the weight matrices used to generate the query, key, and value matrices with $n=1,\ldots,N_{\mathrm{A}}$. The outputs of $N_{\mathrm{A}}$ heads are vertically concatenated as $\mathbf{R}=[\mathbf{r}_{1}^T,\ldots,\mathbf{r}_{N_{\mathrm{A}}}^T]^T$. Then $\mathbf{R}$ is transformed linearly to yield $\mathbf{o}\in \mathbb{R}^{128}$ as the output of the multi-head attention layer. The feature map $\mathbf{o}$ is processed by Dense2 layer with the dimension increased to $256$, followed by the BN and dropout operation. Finally, Dense3 layer outputs the probability vector $\hat{\mathbf{p}}$ by using Softmax activation function. SELU is applied for Bi-LSTM1, Bi-LSTM2, and Dense2 layers.

\section{Training Procedures of C-AMC Framework}

In this section, the offline and online training procedures of the C-AMC framework are developed for different application scenarios. Since SSCNet and MCNet are integrated to fulfill the modulation classification task, they are trained jointly in an end-to-end manner for both of two modes.

\subsection{Offline Training}

\begin{figure}[t]
	\centering
	\includegraphics[width=3.4in]{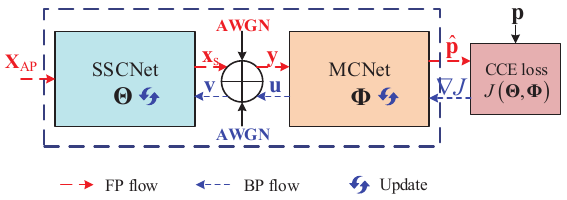}
	\caption{Offline training procedure of C-AMC framework.}\label{C_AMC_offtrain}
\end{figure}

The offline training means to train SSCNet and MCNet centralizedly in the simulation environment and then deploy them separately at the edge device and server. Resorting to the low-cost offline computing resource, this mode can provide a decent base model provided that a certain amount of representative training data are available.

Fig.~\ref{C_AMC_offtrain} illustrates the offline training procedure, where SSCNet and MCNet are connected by the AWGN. Note that the Gaussian distributed noise can be safely used to model the impact of the effective noise $\mathbf{w}$ imposed by the wireless transmission channel since it corresponds to the lower bound of the transmission capacity. Each training sample is paired by the input $\mathbf{X}_{\textrm{AP}}$ and the label $\mathbf{p}$, where $\mathbf{p}$ is a one-hot vector indicating the true modulation type of the source signal. Each time of weight update includes two phases detailed as follows.

\textbf{1) Forward Propagation (FP):} In the FP flow, SSCNet receives the input $\mathbf{X}_{\textrm{AP}}$ and outputs the semantic embedding $\mathbf{x}_{\textrm{s}}$. By adding the generated AWGN $\bar{\mathbf{w}}$ on $\mathbf{x}_{\textrm{s}}$, the noisy semantic embedding can be written as
\begin{eqnarray}
	\label{eqn_y_sim}
	\mathbf{y}=\mathbf{x}_{\textrm{s}}+\bar{\mathbf{w}},
\end{eqnarray}
which acts as the input of MCNet. Then MCNet outputs the predicted probability vector $\hat{\mathbf{p}}$ to approximate $\mathbf{p}$. The categorical cross-entropy (CCE) loss function is used to measure the discrepancy between the prediction and label vectors on a batch of training samples, that is,
\begin{eqnarray}
	\label{eqn_loss}
	J(\boldsymbol{\Theta},\boldsymbol{\Phi}) = -\frac{1}{N_{\textrm{bat}}}\sum_{n=1}^{N_{\textrm{tr}}}\sum_{m=1}^{M} p_{m}^{(n)}\log \hat{p}_{m}^{(n)},
\end{eqnarray}
where $N_{\textrm{bat}}$ denotes the batch size, $p_{m}^{(n)}$ and $\hat{p}_{m}^{(n)}$ respectively represent the $m$th element of $\mathbf{p}$ and $\hat{\mathbf{p}}$ corresponding to the $n$th sample.

\textbf{2) Back Propagation (BP) \& Update:} The error gradient $\nabla J$ is back propagated to SSCNet and MCNet to update $\boldsymbol{\Theta}$ and $\boldsymbol{\Phi}$. In the BP flow, $\boldsymbol{\Phi}$ is directly updated as
\begin{eqnarray}
	\label{eqn_update_Phi}
	\boldsymbol{\Phi}\gets \boldsymbol{\Phi}-\eta\nabla_{\boldsymbol{\Phi}} J,
\end{eqnarray}
where $\eta$ denotes learning rate for the weight update. Based on $\nabla_{\boldsymbol{\Phi}} J$, the error gradient with respect to $\mathbf{x}_{\textrm{s}}$
\begin{eqnarray}
	\label{eqn_gradJ_xs}
	\mathbf{u}=\nabla_{\mathbf{x}_{\textrm{s}}} J=\left[\frac{\partial J}{\partial x_{\textrm{s},1}},\ldots,\frac{\partial J}{\partial x_{\textrm{s},N}}\right]
\end{eqnarray}
is calculated. Adding the AWGN $\bar{\mathbf{w}}^{\prime}$ on $\mathbf{u}$ yields the noisy gradient expressed as
\begin{eqnarray}
	\label{eqn_v_sim}
	\mathbf{v}=\mathbf{u}+\bar{\mathbf{w}}^{\prime},
\end{eqnarray}
which is used to calculate the approximated error gradient with respect to the weights in $\boldsymbol{\Theta}$ as per the chain rule. For an arbitrary weight $\theta$ in $\boldsymbol{\Theta}$, its approximated error gradient is given by

\setlength{\arraycolsep}{0.1em}
\begin{eqnarray}
	\label{eqn_gradJ_theta_approx}
	\hat{\nabla}_{\theta} J&&=\sum_{n=1}^{N} v_{n} \frac{\partial x_{\textrm{s},n}}{\partial \theta}=\sum_{n=1}^{N} \left(\frac{\partial J}{\partial x_{\textrm{s},n}} \frac{\partial x_{\textrm{s},n}}{\partial \theta}+\bar{w}^{\prime}_{n}\frac{\partial x_{\textrm{s},n}}{\partial \theta}\right) \nonumber \\
	&&=\nabla_{\theta} J + \sum_{n=1}^{N} \bar{w}^{\prime}_{n}\frac{\partial x_{\textrm{s},n}}{\partial \theta},
\end{eqnarray}
where $v_{n}$ and $\bar{w}^{\prime}_{n}$ denote the $n$th elements of $\mathbf{v}$ and $\bar{\mathbf{w}}^{\prime}$, respectively. Denoting $\hat{\nabla}_{\boldsymbol{\Theta}} J$ as the set including all the approximated error gradients, $\boldsymbol{\Theta}$ can be updated as
\begin{eqnarray}
	\label{eqn_update_Theta}
	\boldsymbol{\Theta}\gets \boldsymbol{\Theta}-\eta\hat{\nabla}_{\boldsymbol{\Theta}} J.
\end{eqnarray}
The addition of $\bar{\mathbf{w}}$ and $\bar{\mathbf{w}}^{\prime}$ introduces the perturbation in the training stage and thus improves the generalization capability of the C-AMC framework.

\subsection{Online Training}

In some practical scenarios, there are not enough training data available prior to model deployment or the sensed data used for modulation classification exhibit the non-stationary property, in which cases the online training needs to be invoked.

\begin{figure}[t]
	\centering
	\includegraphics[width=3.4in]{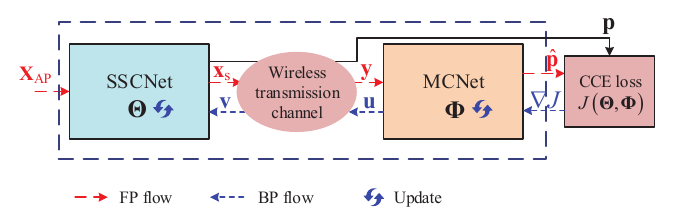}
	\caption{Online training procedure of C-AMC framework.}\label{C_AMC_ontrain}
\end{figure}

Fig.~\ref{C_AMC_ontrain} illustrates the online training procedure. Different from the offline training mode inserting the generated AWGN between SSCNet and MCNet, SSCNet and MCNet are separated by the wireless transmission channel in the online training mode as they have been deployed at the edge device and server, respectively. For the online mode, another main difference from the offline mode is that the training data with form of $\langle\mathbf{X}_{\textrm{AP}}, \mathbf{p}\rangle$ are collected by the edge device sequentially. In the following, the FP and BP \& update phases are briefly described by indicating the detailed differences therein.

\textbf{1) FP:} In the online mode, the noisy semantic embedding acting as the input of MCNet experiences the interface processing and becomes
\begin{eqnarray}
	\label{eqn_y_real}
	\mathbf{y}=\mathbf{x}_{\textrm{s}}+\mathbf{w},
\end{eqnarray}
where $\mathbf{w}$ is the the effective noise compounding AWGN and the interference after the channel equalization. In addition to $\mathbf{x}_{\textrm{s}}$, the edge device also needs to transmit the label $\mathbf{p}$ to the edge server for loss calculation. After accumulating a batch of training samples, the edge server can calculate the CCE loss as per (\ref{eqn_loss}).

\textbf{2) BP \& Update:} Most steps of this phase are same as the counterparts in the offline mode while the difference lies in the noisy gradient. Similar to the form of $\mathbf{y}$, the noisy gradient undergoes the inverse transmission processing and becomes the superposition of $\mathbf{u}$ and the effective noise $\mathbf{w}^{\prime}$, that is,
\begin{eqnarray}
	\label{eqn_v_real}
	\mathbf{v}=\mathbf{u}+\mathbf{w}^{\prime}.
\end{eqnarray}
Then SSCNet and MCNet can be jointly trained by using the sequentially arrived training samples even though they are deployed at two ends of the wireless channel.

It is noted that the offline and online training modes also can be adopted together in practical application. Specifically, the offline training provides the decent base models of SSCNet and MCNet for deployment. After that the online training is executed to adapt the models by using only a few training samples so that the C-AMC framework is able to cope with the dynamics of the spectrum environment and system state.

\section{C-AMC Model Compression}

In this section, the model compression strategy of the C-AMC framework is studied by resorting to weight pruning and quantization, in order to lower the requirements of computation and storage, especially for the resource-constrained edge device. The computational complexity of the C-AMC models is finally analyzed to shed light on the effect of model compression.

\subsection{Weight Pruning}

Weight pruning can reduce the quantity of parameters in a DL model significantly by cutting off neuronal connections with the trivial contribution to the prediction accuracy and then fine-tuning the model, incurring the limited performance loss. The weight magnitude is a metric to measure the weight contribution and those weights with relatively small absolute values will be removed. Specifically, consider a weight $w$ with the small absolute value connecting neuron $i$ and $j$ in a DL model. The output of neuron $i$ and the input of neuron $j$ are denoted by $\alpha$ and $\beta$, respectively. Then we have
\begin{eqnarray}
	\label{eqn_simple_neu_connect}
	\beta=w\alpha+B,
\end{eqnarray}
where $B$ denotes the part of input from other weights for neuron $j$. From the FP flow perspective, small $|w|$ weakens the impact of $\alpha$ on $\beta$, leading to the limited contribution of $w$. From the BP flow perspective, to update the weight $v$ of a connection contributing to the input of neuron $i$, its gradient is calculated as
\begin{eqnarray}
	\label{eqn_grad_v}
	\frac{\partial \mathcal{L}}{\partial v}=\frac{\partial \mathcal{L}}{\partial \beta}\frac{\partial \beta}{\partial \alpha}\frac{\partial \alpha}{\partial v} =\frac{\partial \mathcal{L}}{\partial \beta}w\frac{\partial \alpha}{\partial v},
\end{eqnarray}
where $\mathcal{L}$ denotes the loss function. It can be seen that $w$ with the small absolute value also weakens the gradients of weights related to $w$ in previous layers and thus is less important to the model training. Therefore, the weights with small absolute values contribute trivially to both the model training and inference and the magnitude-based criterion is applied to prune the C-AMC framework.

For the magnitude-based pruning, the pruning threshold is usually determined as per the pruning ratio. Specifically, denote $\mathcal{W}_{i}$ and $\rho_{i}$ as the weight set and pruning ratio of the $i$th layer in the C-AMC framework with $i=1,\ldots,L_{\mathrm{SSC}}+L_{\mathrm{MC}}$, where $L_{\mathrm{SSC}}$ and $L_{\mathrm{MC}}$ represent the numbers of layers in SSCNet and MCNet, respectively. Sort all $N_{\mathcal{W}_{i}}$ weights in $\mathcal{W}_{i}$ in ascending order as per the absolute values to yield the ordered vector $\boldsymbol{\zeta}_{i}$, whose $\lfloor\rho_{i} N_{\mathcal{W}_{i}}\rfloor$th element, $\zeta_{i,\lfloor\rho_{i} N_{\mathcal{W}_{i}}\rfloor}$, is selected as the pruning threshold. Then the $j$th weight in $\mathcal{W}_{i}$ is filtered as
\begin{eqnarray}
	\label{eqn_Prun}
	\mathcal{W}_{i}(j)=
	\begin{cases}
		\mathcal{W}_{i}(j), & \left| \mathcal{W}_{i}(j)\right| \geq \zeta_{i,\lfloor\rho_{i} N_{\mathcal{W}_{i}}\rfloor}\\
		0, & \textrm{otherwise}
	\end{cases},
\end{eqnarray}
for $j=1,\ldots,N_{\mathcal{W}_{i}}$. After pruning all layers of interest in the C-AMC framework, both SSCNet and MCNet are fine-tuned based on the training set to compensate the performance loss.

\subsection{Weight Quantization}

Weight quantization can further reduce the model size and accelerate the model inference, which is a good partner of weight pruning to facilitate the model deployment, especially for the edge device with limited resources.

To reduce the complexity, the post-training quantization is adopted since it is a push-button strategy decoupling the model training and quantization. Consider the $b$-bit uniform affine quantization mapping each reserved weight after pruning to an unsigned integer from the set $\{0,\ldots,2^{b}-1\}$ \cite{M. Nagel}, which is expressed as
\begin{eqnarray}
	\label{eqn_Gen_Quan}
	\mathbb{Q}\left(\mathcal{W}_{i}(j)\right)=\textrm{clamp}\left(\textrm{round}\left(\frac{\mathcal{W}_{i}(j)}{S}\right)+Z;0,2^{b}-1\right),
\end{eqnarray}
where $\textrm{round}(\cdot)$ represents the round-to-nearest operation with $\textrm{clamp}(\cdot;\cdot,\cdot)$ given by
\begin{eqnarray}
	\label{eqn_clamp}
	\textrm{clamp}(x;\mu_1,\mu_2)=
	\begin{cases}
		\mu_1, & x<\mu_1,\\
		x, & \mu_1 \leq x \leq \mu_2,\\
		\mu_2, & x>\mu_2,
	\end{cases}.
\end{eqnarray}
In (\ref{eqn_Gen_Quan}), $S$ and $Z$ denote a scaling factor and an integer zero point quantized from the real zero, which are respectively calculated as
\begin{eqnarray}
	\label{eqn_ScaleFactor}
	S=\frac{\max(\mathcal{W}_{i})-\min(\mathcal{W}_{i})}{2^{b}-1},
\end{eqnarray}
\begin{eqnarray}
	\label{eqn_ZeroPoint}
	Z=-\textrm{round}\left(\frac{(2^{b}-1)\min(\mathcal{W}_{i})}{\max(\mathcal{W}_{i})-\min(\mathcal{W}_{i})}\right),
\end{eqnarray}
where $\max(\mathcal{W}_{i})$ and $\min(\mathcal{W}_{i})$ denote the maximum and minimum values of the weights in $\mathcal{W}_{i}$.  After weight pruning and quantization, the model compression ratios of SSCNet, MCNet, and the C-AMC framework are respectively given by
\begin{eqnarray}
	\label{eqn_CompreRatio_SSC}
	\gamma_{\mathrm{SSC}}=\frac{32}{b}\sum_{l=1}^{L_{\mathrm{SSC}}}\frac{N_{\mathcal{W}_l}}{N_{\mathrm{SSC}}}\frac{1}{1-\rho_l},
\end{eqnarray}
\begin{eqnarray}
	\label{eqn_CompreRatio_MC}
	\gamma_{\mathrm{MC}}=\frac{32}{b}\sum_{l=L_{\mathrm{SSC}}+1}^{L_{\mathrm{SSC}}+L_{\mathrm{MC}}}\frac{N_{\mathcal{W}_l}}{N_{\mathrm{MC}}}\frac{1}{1-\rho_l},
\end{eqnarray}
\begin{eqnarray}
	\label{eqn_CompreRatio_C-AMC}
	\gamma=\frac{N_{\mathrm{SSC}}\gamma_{\mathrm{SSC}}+N_{\mathrm{MC}}\gamma_{\mathrm{MC}}}{N_{\mathrm{SSC}}+N_{\mathrm{MC}}},
\end{eqnarray}
where $N_{\mathrm{SSC}}$ and $N_{\mathrm{MC}}$ denote the numbers of weights of SSCNet and MCNet, respectively.

\subsection{Complexity Analysis}

\begin{table}[t]
	\centering
	\caption{Computational Complexity of SSCNet and MCNet}\label{Complexity}
	\begin{tabular}{c|c|c|c}
		\hline
		& \makecell{Layer \\ type} & \makecell{Layer \\ index} & Complexity \\
		\hline
		\multirow{3}{*}{SSCNet} & Conv1D-1 & $l=1$ & $\mathcal{O}(2\rho_{l}D_{l,1} D_{l,2}N_{\mathcal{W}_l})$ \\
		\cline{2-4}
		& Conv1D-2 & $l=2$ & $\mathcal{O}(2\rho_{l}D_{l,1} D_{l,2}  N_{\mathcal{W}_l})$ \\
		\cline{2-4}
		& Dense1 & $l=3$ & $\mathcal{O}(2\rho_{l}N_{\mathcal{W}_l})$\\
		\hline
		\multirow{6}{*}{MCNet} & Bi-LSTM1 & $l=4$ & \makecell{$\mathcal{O}(4\rho_{l}N( K_{\mathrm{Bin},l} K_{\mathrm{Bout},l}$\\ $ + K^2_{\mathrm{Bout},l}))$} \\
		\cline{2-4}
		& Bi-LSTM2 & $l=5$ &  \makecell{$\mathcal{O}(4\rho_{l}N( K_{\mathrm{Bin},l} K_{\mathrm{Bout},l}$\\ $ + K^2_{\mathrm{Bout},l}))$} \\
		\cline{2-4}
		& \makecell{Multi-head\\ attention} & $l=6$ & \makecell{$\mathcal{O}(\rho_{l}N_{\mathrm{A}}d_{\mathrm{A}}(3K_{\mathrm{Ain}}+ K_{\mathrm{Aout}} ) $\\$+2N_{\mathrm{A}}d_{\mathrm{A}}^2)$} \\
		\cline{2-4}
		& Dense2 & $l=7$ &$\mathcal{O}(2\rho_{l}N_{\mathcal{W}_l})$\\
		\cline{2-4}
		& Dense3 & $l=8$ &$\mathcal{O}(2\rho_{l}N_{\mathcal{W}_l})$\\
		\hline
	\end{tabular}
\end{table}

In this subsection, the model inference complexity of the C-AMC framework is analyzed by using the complexity of floating-point operations (FLOPs) as the metric.

The computational complexities of SSCNet and MCNet are listed in Table~\ref{Complexity} layer by layer. Specifically, $D_{l}$ denotes the length of output feature maps for the Conv1D layer with the index $l$. $K_{\mathrm{Bin},l}$ and $K_{\mathrm{Bout},l}$ denote the input and output dimensions of the unit for the Bi-LSTM layer with the index $l$. $K_{\mathrm{Ain}}$ and $K_{\mathrm{Aout}}$ denote the input and output dimensions for the multi-head attention layer. The number of weights, $N_{\mathcal{W}_{i}}$, is introduced for Conv1D and dense layers to simplify the expression. From Table~\ref{Complexity}, the theoretical complexity is reduced significantly thanks to the weight pruning. If further considering the weight quantization, the model can be more computationally effective with the smaller size.

Then the complexity of the C-AMC framework is given by
\begin{eqnarray}
	\label{eqn_Complexity_CAMC}
	\mathcal{C}_{\mathrm{C\text{-}AMC}}=\mathcal{C}_{\mathrm{SSC}}+\lambda\mathcal{C}_{\mathrm{MC}},
\end{eqnarray}
where $\lambda$ denotes the computation cost factor. Since the computation cost at the edge server is much lower than that in the edge device, $\lambda \ll 1$ usually holds true, leading to a much lower effective complexity for the C-AMC framework.

\section{Simulation Results}

In this section, simulation results are provided to demonstrate the superiority of the proposed C-AMC framework as well as to reveal some insights for the practical design.

\subsection{Simulation Settings}

The widely recognized RadioML2016.10A dataset is used, which includes $11$ modulation types at different signal-to-noise ratios (SNRs). The sampling length of the source signal is set as $L=512$. For the regular training of the C-AMC framework, the training, validation, and testing set contain $132,000$, $44,000$, and $44,000$ samples, respectively. The initial learning rate is $0.001$ and the batch size is $200$. Three baseline schemes listed below are used for performance comparison.

\begin{itemize}[\IEEEsetlabelwidth{Z}]
	\item[\text{-}] \textbf{LSTMNet-DC \cite{S. Rajendran}}. This scheme deploys the LSTM network proposed in \cite{S. Rajendran} at the edge device to classify the modulation types directly without the collaboration of the edge server.
	\item[\text{-}] \textbf{SSCNet-DC}. This scheme deploys SSCNet at the edge device to classify the modulation types directly without the collaboration of the edge server.
	\item[\text{-}] \textbf{SplitAMC \cite{J. Park}}. This scheme splits ResNet-18 with four residual blocks into two parts and deploys them at the edge device and server, respectively, for collaborative AMC.
\end{itemize}
Unless otherwise stated, the SNR in simulation results represents the wireless sensing channel SNR.

\subsection{Results}

\begin{figure}[t]
	\centering
	\includegraphics[width=3.2in]{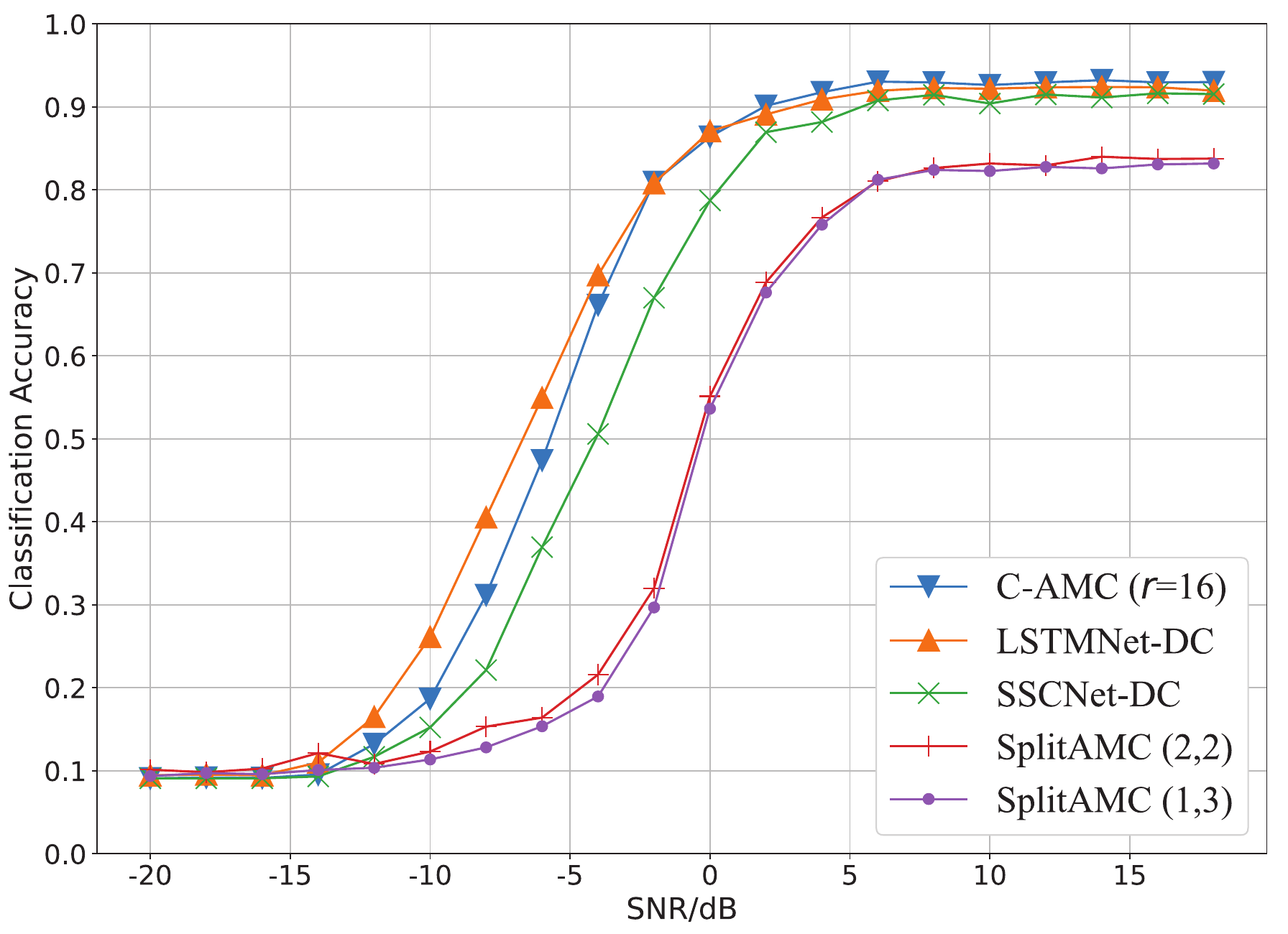}
	\caption{Performance comparison between the C-AMC framework and baseline schemes.}\label{C-AMC_compare_baseline}
\end{figure}

Fig.~\ref{C-AMC_compare_baseline} shows the prediction accuracy versus SNR for the C-AMC framework with $r=16$ and baseline schemes. From Fig.~\ref{C-AMC_compare_baseline}, the C-AMC framework outperforms SplitAMC significantly since the latter relies on a mass of I/Q samples. Compared to SSCNet-DC, the C-AMC framework improves the accuracy substantially thanks to the collaboration of MCNet. LSTMNet-DC achieves the better performance than the C-AMC framework at the low SNR regime, which is not a regime of interest since the prediction accuracy is not high. As the SNR increases, the C-AMC framework gradually surpasses LSTMNet-DC and finally converges at the accuracy over $93$\% with the gain almost $1$\% compared to the latter.
To provide a comprehensive comparison between the C-AMC framework and baseline schemes, Table~\ref{tab:performance_comparison} lists the prediction accuracy, the number of weights, FLOPs, the model inference time, the compression rate $r$, and the security for schemes in Fig.~\ref{C-AMC_compare_baseline}, where the second through forth metrics related to the computational complexity are considered for both the edge device and the edge server. In addition to the accuracy, the C-AMC framework also outperforms SplitAMC in terms of other four metrics, especially the compression rate. Instead of compressing the sampled data, SplitAMC enlarges the dimension of the feature vector and incurs the prohibitively high transmission overhead. The C-AMC framework and SSCNet-DC have the almost same number of weights, FLOPs, and inference time at the edge device, indicating that the performance gain achieved by C-AMC does not incur the additional cost for the edge device. Although the C-AMC framework and LSTMNet-DC have the almost same number of weights, the former incurs fewer FLOPs and thus reduces the inference time by $3\times$ for the edge device yet achieving the higher accuracy. Moreover, the C-AMC framework can protect the secure information by transmitting the intricate semantic embedding instead of the raw data or the classification result. Therefore, the C-AMC framework possesses the unique advantage under the overall consideration on these key metrics.

\begin{table*}[t]
    \centering
	\caption{Comparison of Prediction Accuracy, Weights, FLOPs, and Inference Time}	\label{tab:performance_comparison}
	\begin{tabular}{cccccccccc}
		\toprule
		\makecell{Model} & \makecell{$P_{\mathrm{acc}}$} & \makecell{ Weights\\(Device)} &\makecell{ Weights\\(Server)} & \makecell{FLOPs\\(Device)} & \makecell{FLOPs\\(Server)}& \makecell{Inference Time \\(Device)} &
		\makecell{Inference Time \\(Server)}& \makecell{Compression \\Rate} & \makecell{Security} \\
		\midrule	
		C-AMC ($r=16$)  & \textbf{0.932} & 20.0 K & \textbf{697.0 K} &\textbf{17.9 M} & \textbf{27.4 M} & \textbf{0.022 ms} & \textbf{0.034 ms} & \textbf{16} & $\checkmark$\\
		LSTMNet-DC       & 0.924 & 20.5 K & - & 21.3 M & - & 0.065 ms & - & - & $\times$\\
		SSCNet-DC          & 0.916 & \textbf{18.0 K}& - & \textbf{17.9 M}& - & \textbf{0.022 ms}& - & - & $\times$\\
		SplitAMC(2,2) & 0.839 & 679.7 K& 10.5 M & 150 M & 134 M & 0.133 ms  & 0.107 ms & 0.125 & $\checkmark$\\
		SplitAMC(1,3) & 0.834 & 152.2 K& 11.0 M & 82.5 M & 202.5 M & 0.069 ms & 0.158 ms & 0.0625 & $\checkmark$\\
		\bottomrule
	\end{tabular}
\end{table*}

\begin{figure}[t]
	\centering
	\subfigure[]{
		\includegraphics[width=0.45\linewidth]{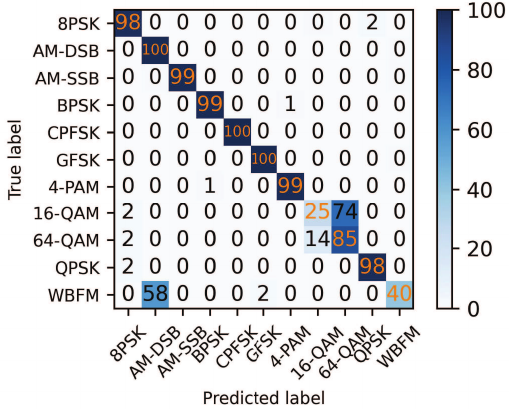}
		\label{SNR=16}
	}
	\subfigure[]{
		\includegraphics[width=0.45\linewidth]{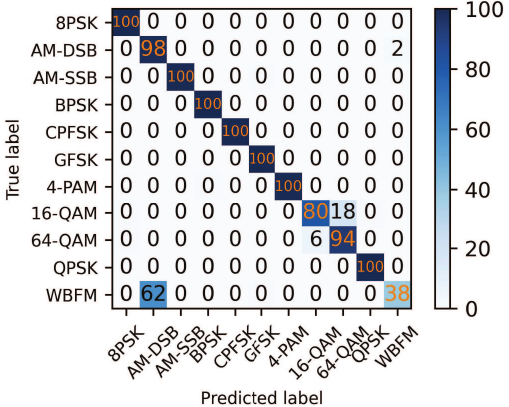}
		\label{SNR=0}
	}
	\vspace{0.5cm}
	
	\caption{The confusion matrices of the C-AMC ($r=16$) at different SNRs: (a) 0 dB, (b) 16 dB.}
	\label{confusion matrices}
\end{figure}

Fig.~\ref{confusion matrices} shows the classification confusion matrices of the C-AMC framework with $r=16$ at $\mathrm{SNR}=0$ dB and $16$ dB, respectively. From Fig.~\ref{confusion matrices}, the accuracy loss of the C-AMC framework mainly comes from 16-QAM, 64-QAM, and WBFM. With the SNR increasing to $16$ dB, the classification accuracy of 16-QAM and 64-QAM improves considerably even with a little confusion between them, leaving WBFM as the dominated destroyer of the overall accuracy.

\begin{figure}[t]
	\centering
	\includegraphics[width=3.2in,height=2.05in]{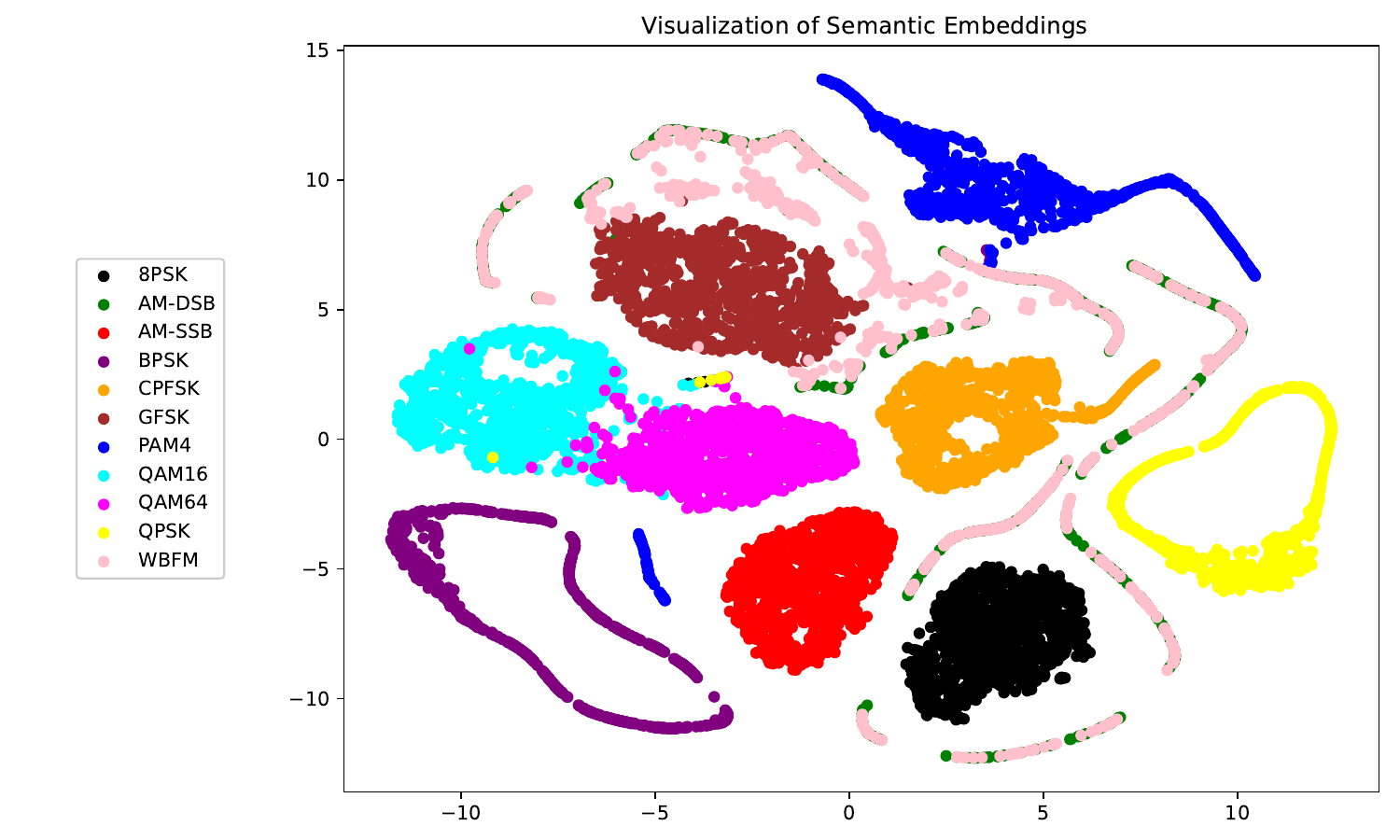}
	\caption{Feature visualization of the semantic embedding with $r=16$.}\label{Visualization}
	
\end{figure}

To reveal the compression effectiveness of SSCNet, Fig.~\ref{Visualization} shows the feature visualization of the semantic embedding $\mathbf{x}_{\textrm{s}}$ with $r=16$. From the figure, the feature points of most modulation types can be clustered and separated from each other despite a few outliers while the feature points of WBFM and AM-DSB are severely overlapped. This phenomenon is coincide with the result in Fig.~\ref{confusion matrices}(b). Therefore, SSCNet can reserve most useful information when compressing the raw data and thus collaborate with MCNet to yield the superior classification accuracy.

\begin{figure}[t]
	\centering
	\includegraphics[width=2.7in]{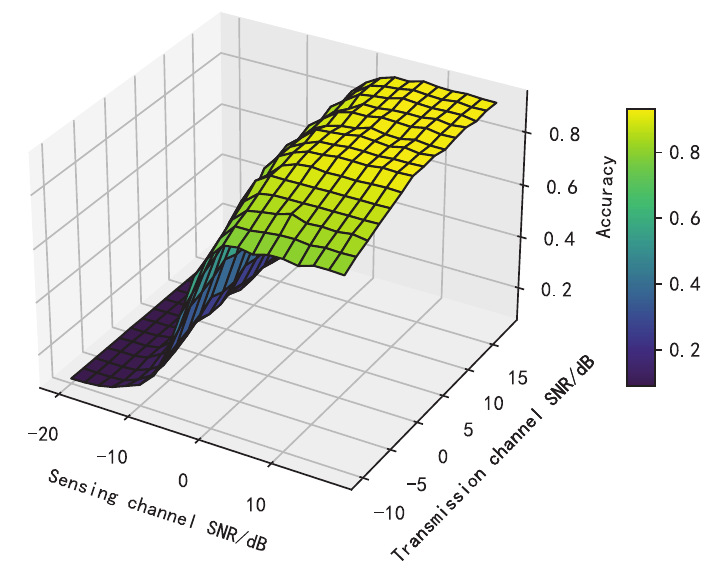}
	\caption{Classification accuracy of C-AMC ($r=16$) versus the sensing and transmission channel SNRs.}\label{C-AMC 3D map}
\end{figure}

Fig.~\ref{C-AMC 3D map} shows the classification accuracy of C-AMC versus the sensing and transmission channel SNRs with $r=16$. From Fig.~\ref{C-AMC 3D map}, the accuracy increases with the sensing channel SNR dramatically while with the transmission channel SNR moderately, revealing that the former dominates the classification performance.

\begin{figure}[t]
	\centering
	\includegraphics[width=3.2in]{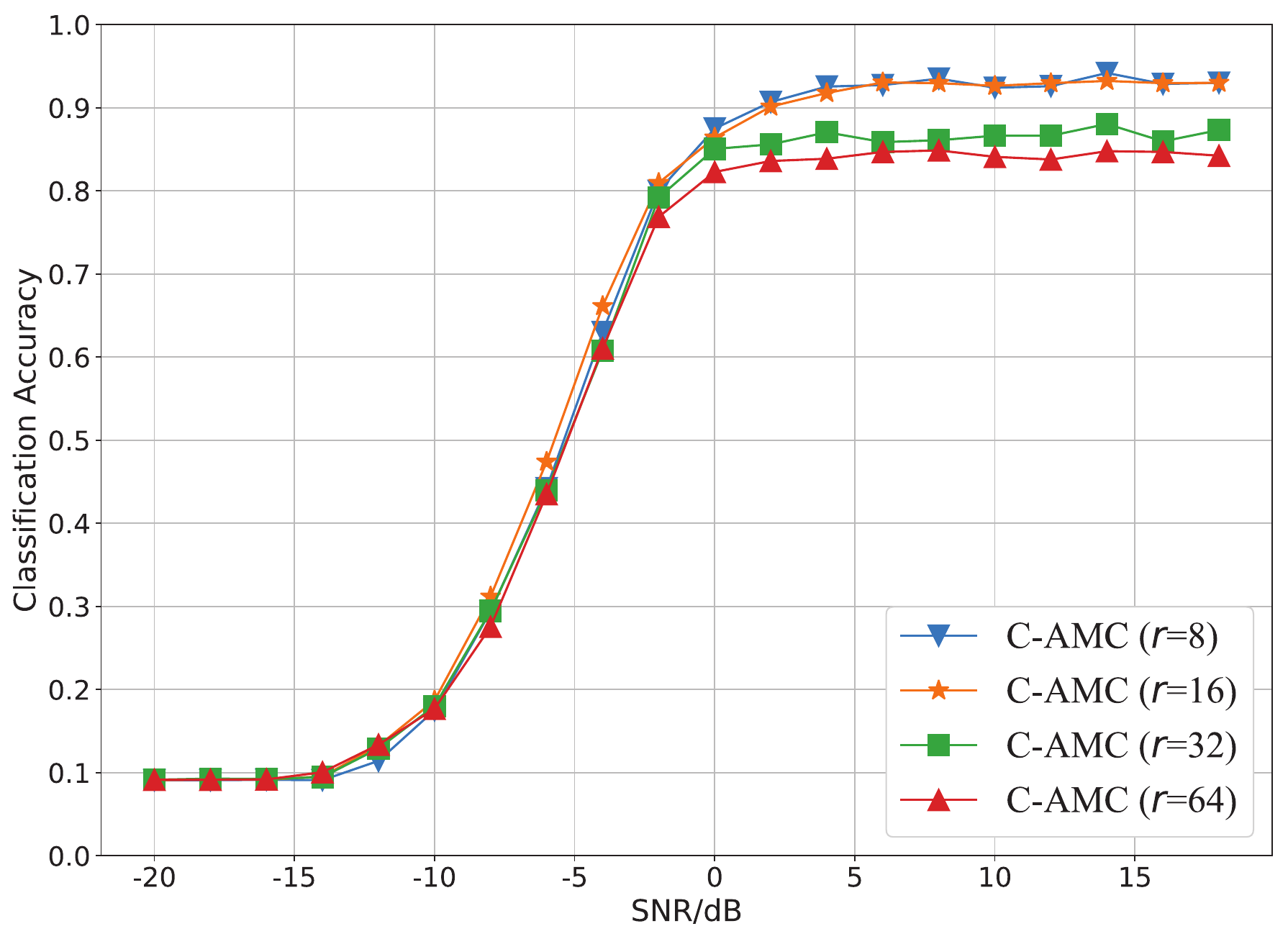}
	\caption{Classification accuracy of C-AMC with different compression rates.}\label{C-AMC compress}
\end{figure}

\begin{figure}[t]
	\centering
	\includegraphics[width=3.2in]{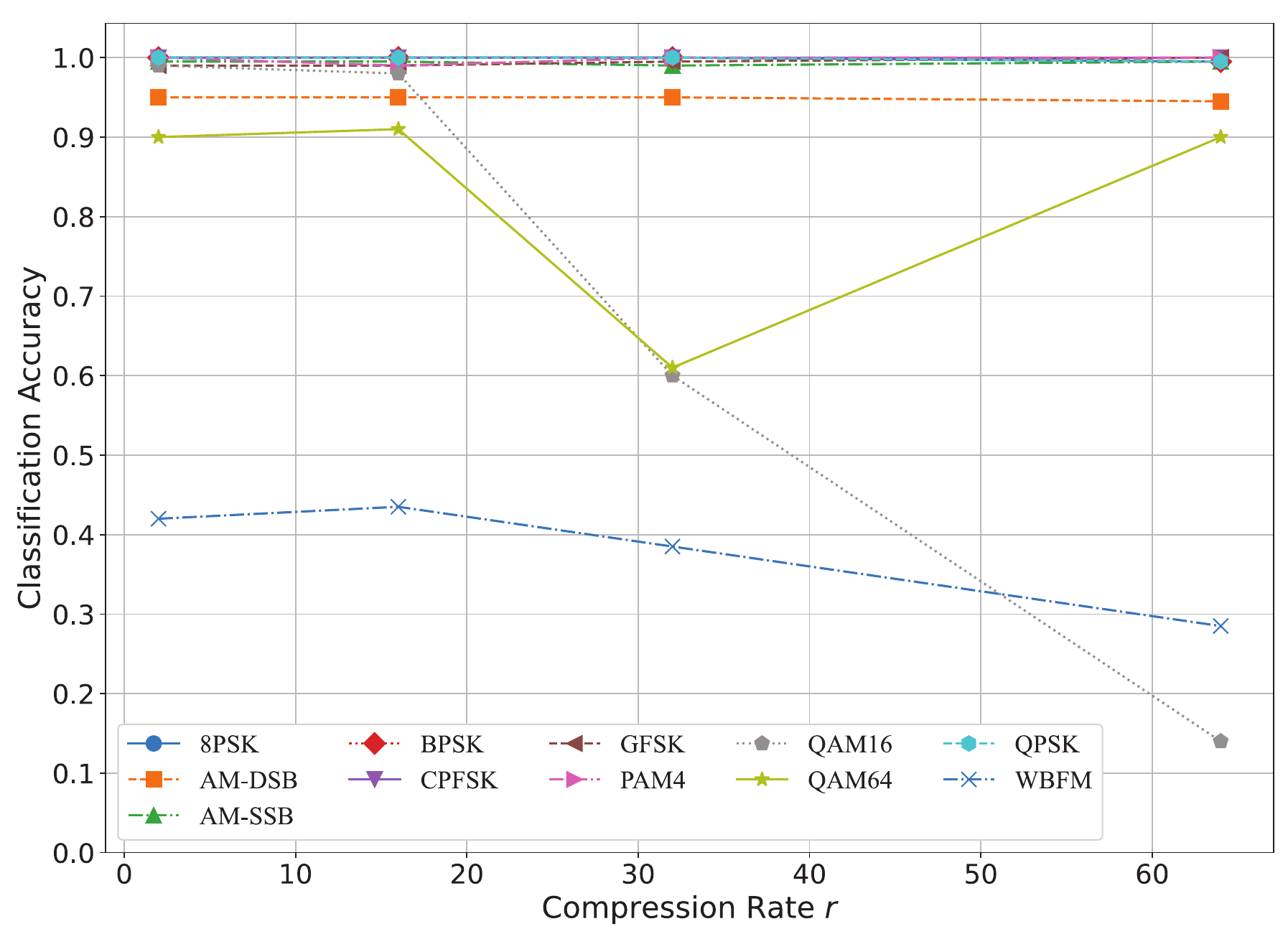}
	\caption{Classification accuracy of C-AMC with different compression rates in terms of the modulation type.}\label{C-AMC compressModu}
\end{figure}

Fig.~\ref{C-AMC compress} shows the classification accuracy of the C-AMC framework with different compression rates. Increasing $r$ from $8$ to $16$ does almost not degrade the accuracy. If the spectrum semantic embedding is compressed with $r=32$, the accuracy decreases nontrivially. The accuracy can be maintained over $80$\% even if $r$ reaches to $64$, demonstrating the robustness of the C-AMC framework to the high compression rate.
To further analyze the reason of performance change versus the compression rate, Fig.~\ref{C-AMC compressModu} shows the accuracy curve of each considered modulation type. From the figure, 16-QAM and 64-QAM are sensitive to the compression along with the confusion between them while WBFM keeps its role of dominated destroyer regardless of $r$. Therefore, if these three modulation types are not considered in some practical applications, the compression rate can reach to at least $64$ without the performance loss.

\begin{figure}[t]
	\centering
	\includegraphics[width=3.2in]{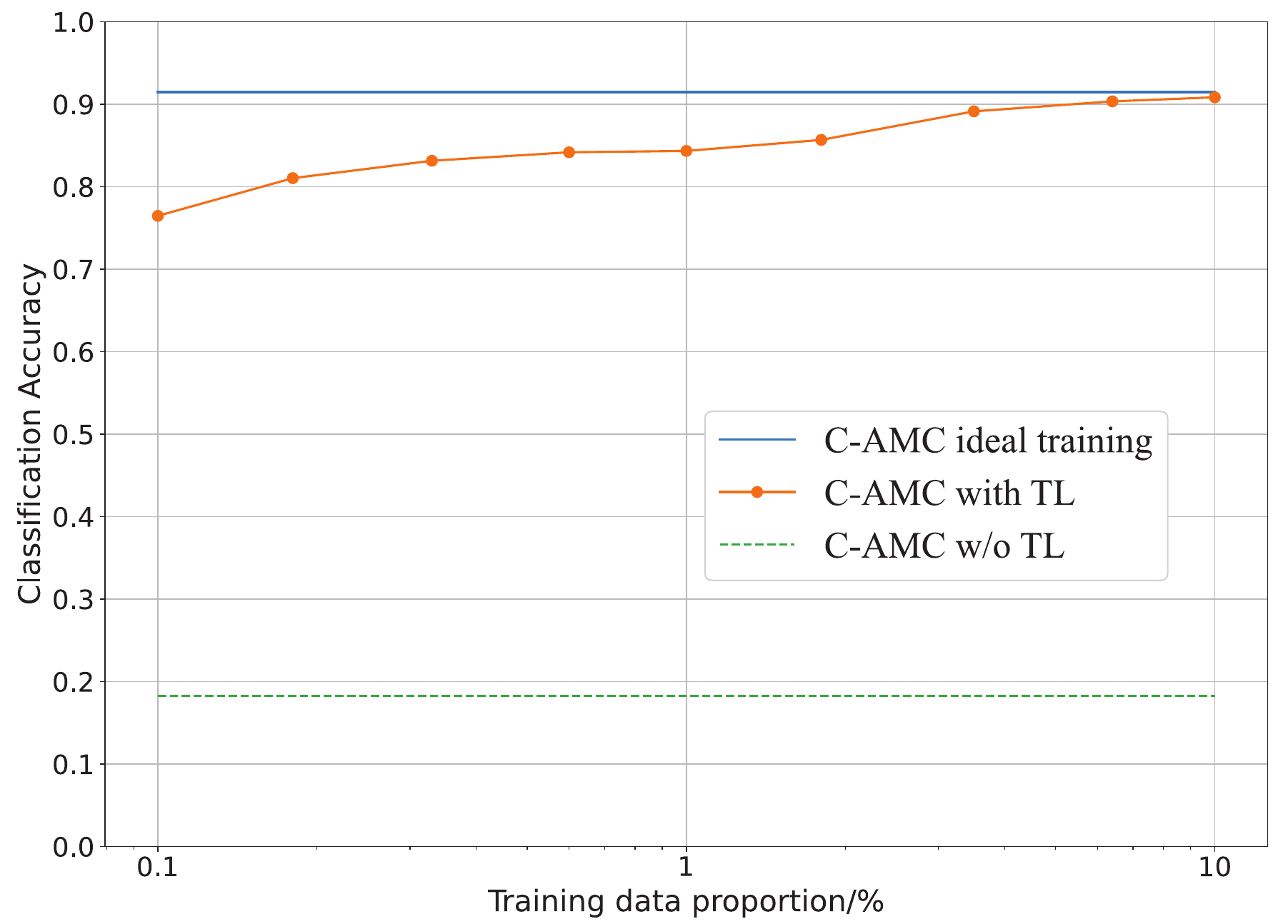}
	\caption{Classification accuracy of C-AMC with the mismatched length of sampling data.}\label{C-AMC TL1}
\end{figure}

\begin{figure}[t]
	\centering
	\includegraphics[width=3.2in]{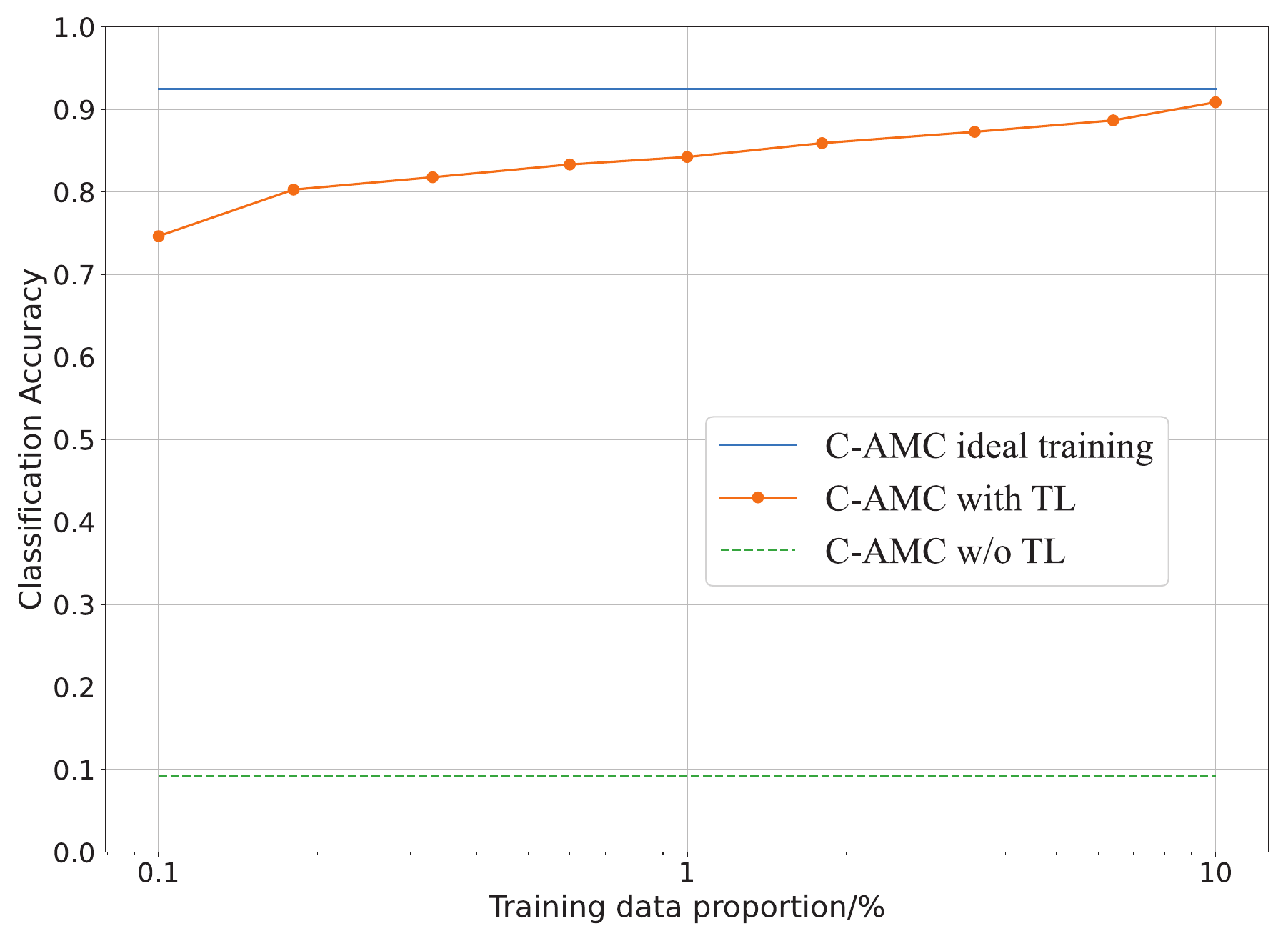}
	\caption{Classification accuracy of C-AMC with the mismatched dataset.}\label{C-AMC TL2}
\end{figure}

Fig.~\ref{C-AMC TL1} shows the classification accuracy of C-AMC with the mismatched length of sampling data, i.e., $L$ is halved. To match the input dimension of SSCNet, $L/2$ zeros are uniformly inserted into the sampled data sequence to yield a new sequence with length $L$. ``C-AMC w/o TL" means directly inputting the new sequence into the C-AMC framework for classification and achieves the very poor performance. ``C-AMC ideal training" means training the C-AMC framework from scratch using sufficient training data with the same format as the new sequence and acts as an upper bound. ``C-AMC with TL" means adapting the original C-AMC framework with a few adaptation data, i.e., transfer learning (TL). The x-axis represents the ratio of the adaptation data quantity to the training data quantity. By resorting to TL, the C-AMC framework can rapidly adapt to the new scenario even with the proportion $0.1$\% and achieve the satisfactory performance with the proportion $1$\%. If the proportion is increased to $10$\%, the accuracy of C-AMC becomes very close to that of ideal training.
Furthermore, Fig.~\ref{C-AMC TL2} shows the classification accuracy of C-AMC with the mismatched dataset, where $11$ different modulation types in the testing set are chosen from RadioML2018 dataset. This figure reveals the similar phenomenon, further validating the robustness of C-AMC when faced with various mismatched cases.

\begin{figure}[t]
	\centering
	\includegraphics[width=3.2in]{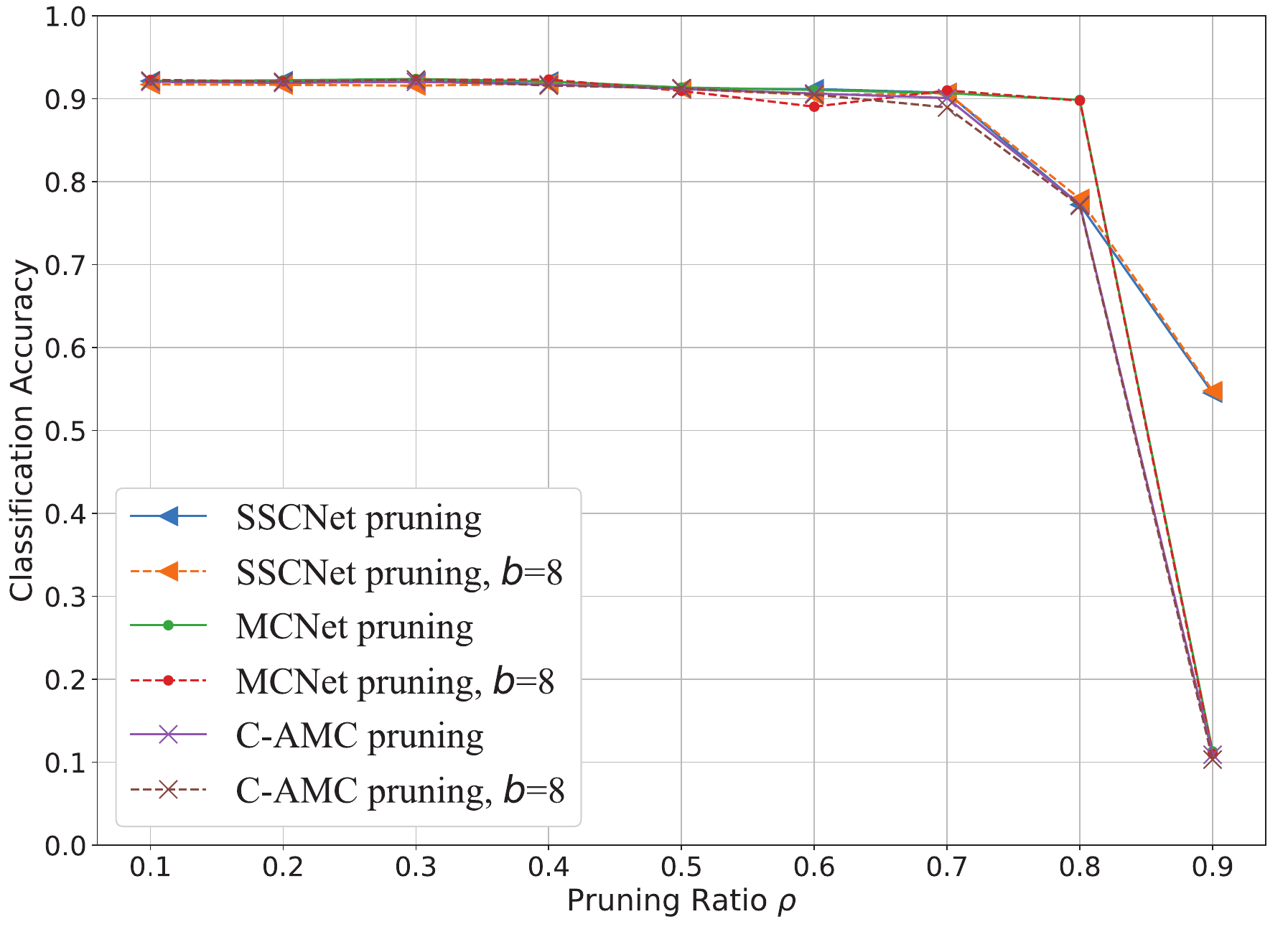}
	\caption{Classification accuracy of C-AMC versus the pruning ratio.}\label{C-AMC_model_compress}
\end{figure}

\begin{table}[t]
	\caption{Comparison of  Weights, FLOPs, and Inference Time for C-AMC}
	\label{tab:performance_comparison1}
	\begin{tabular}{c|c|c|c|c}
		\hline
		\makecell{} & \makecell{Model} &\makecell{ Weights} & \makecell{FLOPs}&
		\makecell{Inference \\Time} \\
		\hline
		\multirow{3}{*}{\makecell{Original}} & SSCNet & 20.0 K & 17.9 M & 0.022 ms \\
		\cline{2-5}
	      	& MCNet & 697.0 K & 27.4 M & 0.034 ms \\
	    \cline{2-5}
	    	& C-AMC & 717.0K & 45.3 M & 0.056 ms \\
	    \hline
		\multirow{3}{*}{\makecell{Pruned}}  & SSCNet($\rho=0.7$) & 6.4 K & 5.7 M & 0.019 ms\\
		\cline{2-5}
		& MCNet($\rho=0.8$) & 147.1 K & 5.7 M & 0.033 ms\\
		\cline{2-5}
		& C-AMC & 153.4 K & 11.4 M & 0.052 ms \\
		\hline
		\multirow{3}{*}{\makecell{Pruned \\and Quantized}} & \makecell{SSCNet\\($\rho=0.7, b=8$)} & 6.4 K & 5.7 M & 0.019 ms \\
		\cline{2-5}
		 & \makecell{MCNet\\($\rho=0.8, b=8$)} & 147.1 K & 5.7 M & 0.032 ms\\
		 \cline{2-5}
		 & C-AMC & 153.4 K & 11.4 M & 0.051 ms \\
		\hline
	\end{tabular}
\end{table}

Finally, Fig.~\ref{C-AMC_model_compress} shows the classification accuracy of C-AMC versus the pruning ratio. Specifically, SSCNet pruning, MCNet pruning, and C-AMC pruning correspond to the following three settings: 1) $\rho_1=\cdots=\rho_{L_{\mathrm{SSC}}}=\rho$, $\rho_{L_{\mathrm{SSC}}+1}=\cdots=\rho_{L_{\mathrm{SSC}}+L_{\mathrm{MC}}}=0$; 2) $\rho_1=\cdots=\rho_{L_{\mathrm{SSC}}}=0$, $\rho_{L_{\mathrm{SSC}}+1}=\cdots=\rho_{L_{\mathrm{SSC}}+L_{\mathrm{MC}}}=\rho$; 3) $\rho_1=\cdots=\rho_{L_{\mathrm{SSC}}+L_{\mathrm{MC}}}=\rho$. From this figure, SSCNet can be pruned up to $\rho=0.7$ with the almost unchanged performance while the threshold for MCNet is $\rho=0.8$. The performance of C-AMC pruning is dominated by the one of SSCNet pruning and MCNet pruning that has the lower accuracy. So the curve of C-AMC pruning coincides with both SSCNet pruning and MCNet pruning with $\rho\leq 0.7$. As $\rho$ increases to $0.8$ and $0.9$, C-AMC pruning respectively coincides with SSCNet pruning and MCNet pruning. In addition, the $8$-bit quantization considered for each pruning strategy does almost not cause the performance loss. Table~\ref{tab:performance_comparison1} lists the number of weights, FLOPs, and the model inference time for the original, pruned, and pruned and quantized C-AMC frameworks. Through the appropriate weight pruning and quantization, i.e., $\rho=0.7$ and $b=8$, SSCNet can be compressed by $\frac{20}{6.4}\times 4\approx 12\times$ with the inference accelerated by $14$\%, further lightening the computation load for the edge device.

\section{Conclusion}

In this article, an EL based C-AMC framework is proposed to meet the requirements of classification accuracy, computation load, transmission overhead, and data privacy for hierarchical CR networks. The C-AMC framework consists of a lightweight SCCNet deployed at the edge device for spectrum semantic compression and an elaborated MCNet deployed at the edge server to predict the modulation type based on the noisy semantic embedding. Both the offline and online training procedures of the C-AMC framework are elaborated. The model pruning and quantization strategy for the C-AMC framework is developed to further facilitate the computation and storage, followed by the comprehensive complexity analysis. Simulation results demonstrate the superiority of the C-AMC framework and reveal useful insights paving the practical implementation.

Compared with the current semantic communication system focusing on semantics of the image, text or speech \cite{H. Xie, Z. Weng}, the spectrum semantics include various concrete types, e.g., the signal modulation type, the emitter type and position, the channel characteristics and the corresponding environment situation. In the future research, it is anticipated to construct a unified framework for the spectrum semantic cognition based on the C-AMC framework.



\ifCLASSOPTIONcaptionsoff
  \newpage
\fi


\begin{thebibliography}{99}
\bibitem{O. A. Dobre}O. A. Dobre, A. Abdi, Y. Bar-Ness, and W. Su, ``Survey of automatic modulation classification techniques: Classical approaches and new trends," \emph{IET Commun.}, vol. 1, no. 2, pp. 137--156, 2007.
\bibitem{P. Panagiotou}P. Panagiotou, A. Anastasopoulos, and A. Polydoros, ``Likelihood ratio tests for modulation classification," in \emph{Proc. 21st Century Mil. Commun. Archit. Technol. Inf. Superiority (MILCOM)}, vol. 2, 2000, pp. 670--674.
\bibitem{Y. A. Eldemerdash}Y. A. Eldemerdash, O. A. Dobre, and M. \"{O}ner, ``Signal identification for multiple-antenna wireless systems: Achievements and challenges," \emph{IEEE Commun. Surv. Tutor.}, vol. 18, no. 3, pp. 1524–1551, Jan. 2016.
\bibitem{B. Dong}B. Dong \emph{et al.}, ``A lightweight decentralized learning-based automatic modulation classification method for resource-constrained edge devices," \emph{IEEE Internet Things J.}, vol. 9, no. 24, pp. 24708--24720, Dec. 2022.
\bibitem{J. L. Xu}J. L. Xu, W. Su, and M. Zhou, ``Likelihood ratio approaches to automatic modulation classification," \emph{IEEE Trans. Syst. Man. Cybern. PartC}, vol. 41, no. 4, pp. 455--469, Jul. 2011.
\bibitem{W. A. Gardner}W. A. Gardner, ``Signal interception: A unifying theoretical framework for feature detection," \emph{IEEE Trans. Commun.}, vol. 36, no. 8, pp. 897--906, Aug. 1988.
\bibitem{D. Boutte}D. Boutte and B. Santhanam, ``A hybrid ICA-SVM approach to continuous phase modulation recognition," \emph{IEEE Signal Proc. Lett.}, vol. 16, no. 5, pp. 402--405, May 2009.
\bibitem{L. Han}L. Han, F. Gao, Z. Li, and O. A. Dobre, ``Low complexity automatic modulation classification based on order-statistics," \emph{IEEE Trans. Wireless Commun.}, vol. 16, no. 1, pp. 400--411, Jan. 2017.
\bibitem{T. J. O'Shea}T. J. O'Shea, J. Corgan, and T. C. Clancy, ``Convolutional radio modulation recognition networks," in \emph{Proc. Int. Conf. Eng. Appl. Neural Netw.}, 2016, pp. 213--226.
\bibitem{T. J. O'Shea_b}T. J. O'Shea, T. Roy, and T. C. Clancy, ``Over-the-air deep learning based radio signal classification," \emph{IEEE J. Sel. Topics Signal Process.}, vol. 12, no. 1, pp. 168--179, Jan. 2018.
\bibitem{S. Peng}S. Peng \emph{et al.}, ``Modulation classification based on signal constellation diagrams and deep learning," \emph{IEEE Trans. Neural Netw. Learn. Syst.}, vol. 30, no. 99, pp. 718--727, Mar. 2018.
\bibitem{B. Tang}B. Tang, Y. Tu, Z. Zhang, and Y. Lin, ``Digital signal modulation classification with data augmentation using generative adversarial nets in cognitive radio networks," \emph{IEEE Access}, vol. 6, pp. 15713--15722, 2018.
\bibitem{F. Meng}F. Meng, P. Chen, L. Wu, and X. Wang, ``Automatic modulation classification: A deep learning enabled approach," \emph{IEEE Trans. Veh. Technol.}, vol. 67, no. 11, pp. 10760--10772, Nov. 2018.
\bibitem{Y. Wang}Y. Wang, M. Liu, J. Yang, and G. Gui, ``Data-driven deep learning for automatic modulation recognition in cognitive radios," \emph{IEEE Trans. Veh. Technol.}, vol. 68, no. 4, pp. 4074--4077, Apr. 2019.
\bibitem{T. Huynh-The}T. Huynh-The, C.-H. Hua, Q.-V. Pham, and D.-S. Kim, ``MCNet: An efficient CNN architecture for robust automatic modulation classification," \emph{IEEE Commun. Lett.}, vol. 24, no. 4, pp. 811--815, Apr. 2020.
\bibitem{Z. Zhang}Z. Zhang, H. Luo, C. Wang, C. Gan, and Y. Xiang, ``Automatic modulation classification using CNN-LSTM based dual-stream structure," \emph{IEEE Trans. Veh. Technol.}, vol. 69, no. 11, pp. 13521--13531, Nov. 2020.
\bibitem{S. Chang}S. Chang, S. Huang, R. Zhang, Z. Feng, and L. Liu, ``Multitask-learning based deep neural network for automatic modulation classification," \emph{IEEE Internet Things J.}, vol. 9, no. 3, pp. 2192--2206, Feb. 2022.
\bibitem{Z. Ke}Z. Ke and H. Vikalo, ``Real-time radio technology and modulation classification via an LSTM auto-encoder," \emph{IEEE Trans. Wireless Commun.}, vol. 21, no. 1, pp. 370–382, Jan. 2022.
\bibitem{H. Zhang}H. Zhang, F. Zhou, Q. Wu, W. Wu, and R. Q. Hu, ``A novel automatic modulation classification scheme based on multi-scale networks," \emph{IEEE Trans. Cognit. Commun. Netw.}, vol. 8, no. 1, pp. 97–-110, 2021.
\bibitem{WangTWC2021}Y. Wang, G. Gui, T. Ohtsuki, and F. Adachi, ``Multi-task learning for generalized automatic modulation classification under non-Gaussian noise with varying SNR conditions,'' \emph{IEEE Trans. Wireless Commu.}, vol. 20, no. 6, pp. 3587--3596, June 2021.
\bibitem{Y. Lin}Y. Lin, Y. Tu, and Z. Dou, ``An improved neural network pruning technology for automatic modulation classification in edge devices," \emph{IEEE Trans. Veh. Technol.}, vol. 69, no. 5, pp. 5703--5706, May 2020.
\bibitem{Z. Chen}Z. Chen \emph{et al.}, "Channel pruning method for signal modulation recognition deep learning models," \emph{IEEE Trans. Cognit. Commun. Netw.}, to be published, 2023.
\bibitem{L. Guo}L. Guo, Y. Wang, Y. Liu, Y. Lin, H. Zhao, and G. Gui, ``Ultralight convolutional neural network for automatic modulation classification in internet of unmanned aerial vehicles,'' \emph{IEEE Internet Things J.}, vol. 11, no. 11, pp. 20831--20839, June 2024.
\bibitem{L. Li}L. Li, J. Huang, Q. Cheng, H. Meng, and Z. Han, ``Automatic modulation recognition: A few-shot learning method based on the capsule network," \emph{IEEE Commun. Lett.}, vol. 10, no. 3, pp. 474--477, Mar. 2021.
\bibitem{W. Lin}W. Lin, D. Hou, J. Huang, L. Li and Z. Han, ``Transfer learning for automatic modulation recognition using a few modulated signal samples," \emph{IEEE Trans. Veh. Technol.}, vol. 72, no. 9, pp. 12391--12395, Sep. 2023.
\bibitem{W. Deng}W. Deng, X. Wang, Z. Huang, and Q. Xu, ``Modulation classifier: A few-shot learning semi-supervised method based on multi-modal information and domain adversarial network," \emph{IEEE Commun. Lett.}, vol. 27, no. 2, pp. 576–-580, Feb. 2023.
\bibitem{P. Qi}P. Qi, X. Zhou, Y. Ding, Z. Zhang, S. Zheng, and Z. Li, ``FedBKD: Heterogenous federated learning via bidirectional knowledge distillation for modulation classification in IoT-edge system," \emph{IEEE J. Sel. Topics Signal Process.}, vol. 17, no. 1, pp. 189--204, Jan. 2023.
\bibitem{WangTCCN2022}Y. Wang, G. Gui, H. Gacanin, B. Adebisi, H. Sari, and F. Adachi, ``Federated learning for automatic modulation classification under class imbalance and varying noise condition,'' \emph{IEEE Trans. Cogn. Commun. Netw}., vol. 8, no. 1, pp. 86--96, March 2022.
\bibitem{S. Zhang}S. Zhang, Y. Yang, Z. Zhou, Z. Sun, and Y. Lin, ``DIBAD: A Disentangled Information Bottleneck Adversarial Defense Method Using Hilbert-Schmidt
  Independence Criterion for Spectrum Security,'' \emph{IEEE Trans. Inf. Forensics Secur.}, vol. 19, pp. 3879--3891, 2024.
\bibitem{P. Dong}P. Dong, Q. Wu, X. Zhang and G. Ding, ``Edge semantic cognitive intelligence for 6G networks: Novel theoretical models, enabling framework, and typical applications," \emph{China Commun.}, vol. 19, no. 8, pp. 1--14, Aug. 2022.
\bibitem{S. Rajendran}S. Rajendran, W. Meert, D. Giustiniano, V. Lenders, and S. Pollin, ``Deep learning models for wireless signal classification with distributed low-cost spectrum sensors," \emph{IEEE Trans. Cognit. Commun. Netw.}, vol. 4, no. 3, pp. 433--445, May 2018.
\bibitem{W. Xu}W. Xu \emph{et al.}, ``Edge learning for B5G networks with distributed signal processing: Semantic communication, edge computing, and wireless sensing," \emph{IEEE J. Sel. Topics Signal Process.}, vol. 17, no. 1, pp. 9--39, Jan. 2023..
\bibitem{J. Park}J. Park, S. Oh and S. -L. Kim, ``SplitAMC: Split learning for robust automatic modulation classification," in \emph{Proc. IEEE 97th Veh. Technol. Conf. (VTC-Spring)}, Florence, Italy, 2023, pp. 1--6.
\bibitem{M. Nagel}M. Nagel \emph{et al.}, ``A white paper on neural network quantization," \emph{arXiv preprint arXiv: 2106.0829}, 2021.
\bibitem{H. Xie}H. Xie, Z. Qin, G. Y. Li, and B.-H. Juang, ``Deep learning enabled semantic communication systems," \emph{IEEE Trans. Signal Process.}, vol. 69, pp. 2663--2675, Apr. 2021.
\bibitem{Z. Weng}Z. Weng and Z. Qin, ``Semantic communication systems for speech transmission," \emph{IEEE J. Sel. Areas Commun.}, vol. 39, no. 8, pp. 2434--2444, Aug. 2021.




\end{thebibliography}
\end{document}